# Development of Rubrics for Capstone Project Courses: Perspectives from Teachers and Students

Rex P. Bringula
College of Computer Studies and
Systems
University of the East
rex_bringula@yahoo.com

## ABSTRACT
This study attempted to develop fair, relevant, and content-valid assessment tools for capstone project courses. Toward this goal, new rating instruments based on the concept of rubrics were proposed. To ensure that the new instruments were valid and fair, several meetings with faculty and students of the computing science departments (i.e., Computer Science and Information Technology) were successively conducted. Eight faculty members and 10 students participated in the study. The final versions of the instruments were completed after a series of careful deliberations with faculty and students. Faculty and students perceived the new instruments fairer than the previous ones. Since the final instruments will be deployed this semester, their strengths and weaknesses are not yet known at this time. Directions for future research are presented.

## Categories and Subject Descriptors
K.3.2 [**Computer and Information Science Education**]: Information systems education

## General Terms
Measurement, Documentation

## Keywords
Assessment Tool, Capstone Project, Rubric, Thesis Writing

## 1. INTRODUCTION
Capstone project courses are one of the key courses in the computing degree programs. Moshkovich [12] argued that computing degree programs (e.g., Computer Science (CS), Information Technology (IT), and Information Systems (IS)) curricula must have courses that could provide opportunities for students to synthesize and apply the knowledge and skills they acquired over several years of study. These courses serve as responses of the academic community to supply the needs of the college graduates who are highly competent technically and who also possess good communication traits, strong leadership abilities, skills as effective team players, and desirable work ethic [10,13].

Miles and Kelm [11] opined that, at the end of capstone courses, students must be aware of the ethical implications of software development, understand social interactions and motivations in customer relations, learn to work effectively with colleagues in a team environment, demonstrate advanced critical thinking skills (e.g., assess the feasibility of the project, analyze the requirements of the software, identify alternative implementation strategies), demonstrate good communication and presentation skills, show project management skills, and understand the dynamics of the human-computer interface. These courses also provide good opportunities for students to apply the knowledge they learned from previous courses, develop communication skills, demonstrate problem-solving skills, and gain firsthand information as to how knowledge is produced [5,8,17].

In the Philippines, course offerings are mandated by the Commission of Higher Education (CHED). CHED [4] issued memoranda of minimum requirements, standards and policies with regard to the Philippine higher education. It agrees with Lunt et al. [9] that capstone courses are usually offered for one or two semesters during the final year of the students. To differentiate between the capstone projects of CS and IT degree programs, the former are required to take Thesis Writing while the latter are required to do the Capstone Project. Though the names of the courses are different, they also share the same curriculum principle, i.e., students are given opportunities to apply their skills and knowledge in solving challenging problems [9].

It cannot be doubted that capstone courses are one of the difficult, demanding, and challenging courses in the computing curricula [2]. The practice of requiring the students to defend their projects in a panel consisting of three members adds complexity and difficulty in meeting the requirements of the course. Students invest money, time, and effort in order to pass it. A delay of one or two semesters on the completion of the capstone projects due to failure in the oral defense is translated to additional expenses for the students and their parents. Consequently, they could not graduate at the expected time.

The pressing concerns of fairness and the introduction of outcomes-based education in the Philippines [3] warrant the need to develop fair, relevant, and valid assessment instruments. Specifically, the study aimed 1) to report the steps undertaken in the development of capstone project rubrics, and 2) to present and discuss the capstone project rubrics.

## 2. LITERATURE REVIEW
Beyerlein et al. [1] proposed a framework in developing an assessment instrument in capstone engineering design courses. It incorporated the perspectives of the educational researcher, the student learner, and the professional practitioner. To this aim, the researchers identified the performance areas for engineering design. The performance areas were personal capacity, team processes, solution requirements, and solution assets. Personal capacity refers to the individual performance and skills improvement on engineering design. Team processes involved the development and implementation of collective processes that supported team design productivity. The third performance area (i.e., solution requirements) defined the goal state for design activities and features expected as required by stakeholders' needs and constraints. The last performance area (i.e., solution assets) referred to the results from a design project that meets the needs and satisfaction of the stakeholders. The researchers further commented that rubrics were the tools that could help capstone instructors to measure higher-level conceptual knowledge, performance skills, and attitudes.

Meyer [10] identified the common learning outcomes of all Electronics and Communications Engineering (ECE) capstone design outcomes at Purdue University. According to the author, the learning outcomes of ECE capstone design outcomes were 1) an ability to apply knowledge obtained in earlier course and to obtain new knowledge necessary to design and test a system, component, or process to meet desired needs, 2) an understanding of the engineering design process, 3) an ability to function on a multidisciplinary team, 4) an awareness of professional and ethical responsibility, and 5) an ability to communicate effectively, in both oral and written form. Outcomes 2 and 4 were measured using rubrics. For Outcome 2, written reports were assessed in terms of their technical content, update record/completeness, professionalism, and clarity/organization. A score of 0-10 can be given to the components and each component had different weights. Technical content and professionalism each had a weight of 3 points while Update record/Completeness and Clarity/Organization had weights of 2 points each. Meanwhile, Outcome 4 was assessed in terms of Introduction, Results of Patent Search, Analysis of Patent Liability, Action Recommended, List of References, and Technical Writing Style. The criteria can also be rated using 0-10 but had different weights. Introduction, Action Recommended, List of References, and Technical Writing Style had weight of 1 point each while Results of Patent Search and Analysis of Patent Liability had weight of 3 points each.

The study of Pauzi and Muda [12] described the assessment of Capstone Civil Engineering Design students in Universiti Tenaga Nasional (UNITEN). They reported that the works of the students were assessed in terms of written reports (20%), conceptual and detailed design (25%), formal presentations (25%), tender documents with the construction cost estimates, and project participation and team works. It made use of rubrics with 7 criteria to assess the students' work in terms of their teamwork and participation in Capstone Design Project course. The seven criteria were Workload (share of task), Getting Organized (initiative to conduct a meeting and make the group organize), Participation in Presentation (participation in sharing ideas, feelings, and thoughts), Client Consultancy Meeting Deadline (ability to do tasks on time or ahead of time), Showing up for Meetings (showing up in a meeting punctually or even ahead of time), Providing Feedback on the Comment from the Meeting (participate actively during a meeting), and Receiving Feedback (manner of receiving feedback). Students can have a mark of 3 (minimum) to 20 (maximum) on each criterion.

Rubric was also employed at Stevens Institute of Technology in its systems engineering (SE) framework for multidisciplinary capstone design courses. Sheppard et al. [18] solicited inputs from systems engineering faculty members with extensive industrial experience in the SE field. In general, evaluators assessed the SE capstone of the students in terms of the project and of students' individual contribution. Learning goals and performance criteria were identified on each criterion. Project assessment and individual assessment had five and two learning goals, respectively. The level of achievement on each learning goal was evaluated using the rating points of 1 (poor) to 5 (excellent). It is interesting to note that students could evaluate their teammates and themselves in their contribution on the project. Using the scores 1 (below expectations), 2 (marginal), and 3 (meets or exceeds expectations), they evaluated the team members in terms of contributions of time, effort, and technical expertise, cooperation with other team members, timely completion of individual assignments, and overall contribution to the team. Lee and Lai [7] also advocated the inclusion of team members' participation in order to increase fairness in assessment.

## 3. DEVELOPMENT OF THE INSTRUMENT
### 3.1 Research Locale and Nature of Capstone Project Courses

The study was conducted in the College of Computer Studies and Systems (CCSS) of the University of the East in Manila. CCSS offers five bachelor's degree programs – Computer Science (CS), Information Technology (IT), Information Systems (IS), Digital Animation major in Gaming (DAG), and Digital Animation major in Animation (DAA). Table 1 shows the flow of capstone project courses for these programs. All except DAG undergo Methods of Research for Information Technology (MERIT). MERIT courses are tailored to each degree program. At the end of this course, CS students are expected to come up with a research topic. Once they have defended it successfully, they will further develop the concept in Thesis Writing A. In this stage, the first three chapters of the paper (i.e., Introduction, Literature Review, and Methodology) are completed. The last stage capstone project course for CS students is Thesis Writing B where the software and the full chapters of the paper are presented.

**Table 1. Flow of Capstone Project Courses for Each Degree Program**

| Degree Programs | Sequence of Capstone Courses |
|---|---|
| CS | MERIT → Thesis Writing A → Thesis Writing B |
| IT/IS/DAA | MERIT → Capstone A → Capstone B |
| DAG | Capstone A → Capstone B |

Meanwhile, DAA students follow the same procedure with the CS students except that the course codes were changed. The first three chapters and the software are expected to be presented in Capstone A and Capstone B, respectively. IT/IS students have exactly the same sequence of capstone courses. They will enroll MERIT, Capstone A, and Capstone B. However, the natures of these courses are different from CS and DAA. For IT/IS, MERIT entails submissions of the first three chapters of the paper. Then, provided they pass MERIT, they will enroll Capstone A which requires them to furnish the full paper and the software. Finally, they will implement the software in their client's company during Capstone B.

On the other hand, DAG students have a shortened flow of capstone project. The nature of Capstone A for DAG includes the proposal and the first chapters of project. Then, upon completion of this course, they have to develop their project and implement it at the same time. This is the Capstone B. They all have to comply all of these activities within the span of five semesters.

**Table 2. Intended Learning Outcomes of the Courses**

| Courses | Degree Program | Intended Learning Outcomes of the Courses |
|---|---|---|
| MERIT | CS and IT | • Develop a paper that will exhibit their scholarly manner of writing.<br>• Explain the concept of the paper in a panel consisting of three members.<br>• Demonstrate mastery of communication skills. |
| Thesis Writing A | CS | • Develop further the paper that will exhibit their scholarly manner of writing.<br>• Demonstrate mastery of communication skills. |
| Capstone A | DAA | |
| Thesis Writing B | CS | • Develop the concept of the paper into a working program.<br>• Demonstrate mastery of communication skills. |
| Capstone A | IT/IS/DAA | |
| Capstone A and Capstone B | DAG | (under construction) |
| Capstone B | IT/IS | • Formulate solutions and alternatives when confronted with barriers in system implementation.<br>• Demonstrate teamwork while implementing the system. |

Along with the flow of the capstone project courses, the intended learning outcomes (ILOs) of the courses of each degree program is presented (See Table 2.). Currently, all except capstone project courses of DAG have ILOs. (The ILOs of DAG are still under construction when this paper is being written.). It can be noticed in Table 2 that there are ILOs that are common across courses and students. The demonstration of mastery of communication skills is present in MERIT, Thesis Writing A, Thesis Writing B, and Capstone A. This is because these courses intend to hone the English skills of the students since English is the second language of Filipinos.

It can also be observed that the presentation of the program in Thesis Writing B and Capstone A for IT/IS/DAA is one of the ILOs for these courses. Furthermore, since Capstone B of IT/IS is about system implementation, oral defense is no longer necessary. Thus, it can be concluded that based on the nature of ILOs and of the courses, only two assessment instruments are needed – one for the research proposal and another one for the software. The construction of the first version of the assessment instruments is discussed in the next section.

The college has existing capstone courses assessment tools. There are about five assessment tools. The assessment tools are rating forms that gauged students' oral defense performance and/or paper from 74% and below (fail) and 75% to 100% (passing). A sample assessment tool is given below. As shown in Figure 1, an evaluator can assess the software as well as the paper of the students.

Evaluators had the impression that students have to be gauged using the verbal ratings (e.g., Excellent, Outstanding, Very Good, etc.) on each box. Hence, they wrote "Excellent", "Outstanding", etc. on the boxes. Students, on the other hand, also commented numerical ratings needed to be more descriptive, that is, each numerical rating must compensate their efforts exerted and not as perceived by the evaluator. For example, a grade of 75% for a paper only reflects the evaluator's perceived applicable grade but it does not reflect why such rating was applicable. Further, the recent educational paradigm shift of the University into outcomes-based education intensifies the need to change the assessment tool.

It is proposed that a new assessment tool be developed in a form of rubric. Rubric was proposed because of its perceived benefits. It clearly communicates to students the requirements of the course [13,19]. Also, the assessment becomes clearer, easier, more objective, and sometimes faster [19]. There are only two proposed rubrics as mentioned earlier. For purposes of clarity, proposal stage is composed of courses that require submission of a paper while software project stage is composed of those courses that involve the full paper and the software. Proposal and Software Project measurement criteria are shown in Table 3.

```
UNIVERSITY OF THE EAST
COLLEGE OF COMPUTER STUDIES AND SYSTEMS
FINAL DEFENSE EVALUATION FORM (CPROP / CPP111 – IT)

TITLE OF STUDY: (type your title below)
_______________________________________________

LEGEND: (Note: The highest group grade for (re)defense is 79.)
Excellent – 98-100    Outstanding – 95-97    Very Good – 89-94    Good – 83-88
Fair – 77-82          Low Passed – 75-76     Failed – below 75

I.   CONTENT...........................................................................
        • Representation of actual data is utilized.
        • Appropriate database schema is evident.
        • Dataset is sufficient to complete processing cycles.
        • Necessary reports are generated.
II.  FUNCTIONALITY.................................................................
        • Software is user-friendly.
        • Modules perform the intended purpose.
        • Software possesses necessary modules for a complete processing cycle.
        • Software generates desired results.
III. RELIABILITY.......................................................................
        • Appropriate validation procedures are used.
        • Accuracy of performance is apparent.
        • System failures are tolerable.
        • Software manifests recovery procedure.
IV.  MAINTAINABILITY..............................................................
        • Item management (add, edit, delete, etc.) is present.
        • Provisions for diagnostic tools and procedures exist.
        • Provisions for enhancements and modifications exist.
V.   PAPER..............................................................................
        • The presentation of results is clear.
        • The results of the study show comprehensive analysis and interpretation.
        • Conclusions are logical and recommendations are appropriate.
VI.  PRESENTATION...................................................................
        • Confidence is evident during the oral examination.
        • Demonstrates mastery of the subject matter.
        • Answers questions courteously.
        • Observes appropriate attire.
```

**Figure 1. Sample Evaluation Form**

**Table 3. Proposal and Software Project Criteria of Measurement**

| Proposal Stage | Software Project Stage |
|---|---|
| *Analysis* – It refers to the clarity of discussion of the problem/objectives of the study. | *Functionality* – The ability of software to carry out the functions as specified or desired. |
| *Relevance*<br>• For IT/IS proposals, it refers to the strength of evidence/proof that the proposed software is useful for the company.<br>• For CS/DAA proposals, it refers to the likelihood that the proposal could contribute to the existing body of knowledge. | *Completeness* – A software characteristic wherein it contains all of its necessary modules. |
| *Method* - It signifies the appropriateness of the methods that would be employed to meet the objectives of the study. | *Reliability* – The ability of software to perform a required function under stated conditions for a stated period of time without any errors. |
| *Paper* - It measures the comprehensiveness of the discussion of the paper citing relevant studies. | |
| *Mastery of the Subject* – It evaluates the correctness of the response of the student on evaluators' query. | |

Students under both stages are evaluated in terms of their paper and mastery of the subject matter. During the proposal stage, presentation of analysis, relevance, and methods that will be employed are evaluated. In Analysis, it intends to measure clarity of the discussion of the paper. Students must provide a vivid discussion on how the study has been built. Along with this, they have to show the importance of doing the project in the point of the client or of the academic/scientific community. The methods will also be scrutinized. These criteria have been selected since all courses employ all of these principles while writing a proposal.

Meanwhile, only three software quality criteria were selected. These were selected since all software applications developed in the college for the last two years could be measured using these criteria. Functionality is a software criterion that intends to measure the conformance of the behavior of the software to its expected behavior. The software that should be presented to the evaluator should not lack essential modules (i.e., Completeness). Lastly, the program must be bug-free; hence, it should be reliable.

The points are scaled from 1 (lowest) to 6 (highest) and the highest score that a student can get is 30 points (5 items x 6 points). The six-point scale was selected because the transmuted points would cover all grade points in the University grading scales. Further, the difference of a point in the scale would not make a big leap of points when transmuted. For example, a 28- and 29-total point ratings would be transmuted to 97% (28/30*50+50) and 98, respectively. The 97% and 98% percentage ratings would be then equivalent to 1.25 and 1.00, respectively. Thus, the rating is deemed fair. The first versions of the rubrics are shown in Table 4 and Table 5.

**Table 4. First Version of Proposed Rubric for Proposal Stage**

| Criteria | Guide Items (points) |
|---|---|
| Analysis | The discussion of the problem is **very clear** AND provided **alternative** ideas. (6 points). The discussion of the problem/objective is **very clear (5) / clear (4) / somewhat clear (3) / confusing (2) / very confusing (1)**. |
| Relevance-IT/IS | The evidence is **exceptionally outstanding (6) / very strong (5) / strong (4) / moderately compelling (3) / weak (2) / very weak (1)** that the proposal is useful for the company. |
| Relevance-CS/DAA | The proposal is **exceptionally outstanding (6)** that it would / is **very likely (5)** that it would/ **likely (4) / will probably (3)** that it would / is **unlikely (2)** / is **very unlikely (1)** it would contribute to the academic/ scientific community. |
| Method | The method is **highly appropriate and innovative (6) / highly appropriate (5)/ appropriate (4) / appropriate to a least extent (3) / inappropriate (2) / highly inappropriate (1)**. |
| Paper | Exemplary! It does not only have complete paper, it **surpassed more than what is required (6)**. A comprehensive, clear, and extensive research is **highly evident (5) / evident (4) / moderately evident (3) / evident to a little extent (2) / not evident (1)**. |
| Mastery of the Subject | Proponents do not only answer **all questions correctly** but also **challenge previous concepts** (6). **All (5) / Most (4) / Half (3) / Only few (2) / No (1)** questions were answered correctly. |

**Table 5. First Version of Proposed Assessment Tool for Software Project Stage**

| CRITERIA | |
|---|---|
| Functionality | **Exemplary**! It does not only perform all of its functions, it presented a **novel** way to perform a function (6). / The software was able to perform **all (5) / most (4) / at least half (3) / few (2) / none (1)** of its functions. |
| Completeness | **Exemplary**! It does not only have complete modules, it **surpassed more than what is required!** (6) / The program contains all (5) essential modules. / Exceeds minimum requirements but not 100% completed (4) / Minimum requirements are met (3). / Only a portion of the minimum requirements are satisfied (2). / None of the minimum requirements are met (1). |
| Reliability | The codes are not only bug-free but also written **efficiently** and **elegantly** (6). / The software is bug-free (5). / There are bugs but do not compromise software performance (4). / There are bugs that compromise software performance to **some extent** (3). / There are bugs that **compromise** software performance (2). / There are so many bugs to the extent that the software no longer performs its functions (1). |
| Paper | **Exemplary**! It does not only have complete paper, it **surpassed more than what is required!** (6) / A comprehensive, clear, and extensive research **is highly evident (5) / evident (4) / moderately evident (3) / evident to a little extent (2) / not evident (1)**. |
| Mastery of the Subject | Proponents do not only answer **all questions correctly** but also **challenge previous concepts** (6). **All (5) / Most (4) / Half (3) / Few (2) / No (1)** questions were answered correctly. |

## 3.2 Presentation of Rubrics to Faculty and Students for Comments and Suggestions

The proposed rubrics were presented to seven faculty members of the computing departments. The group was composed of four chairpersons, one representative from the Office of Curriculum Development and Instruction, and two thesis coordinators. They scrutinized the contents of the initial assessment tools. Their comments, suggestions, or concerns are shown in Table 6. There were two sessions of deliberations.

Faculty and students had common as well as distinct concerns on the rubrics. It was found out that faculty and students had a common concern that it was difficult to achieve a rating of 6. However, they noted that the proposed rubrics were fairer than the previous ones. They argued that while it could be difficult to have a point of 6, it is now easier to get at least half of the perfect points. In short, it is difficult to get perfect points but it is easier to pass with the new assessment tools.

**Table 6. Final Version of Proposed Rubric for Proposal Stage**

| Criteria | Guide Item (points) |
|---|---|
| Analysis | The discussion of the problem/objective is **very clear** AND it provides **alternative** ideas. (6 points) The discussion of the problem/objective is **very clear (5) / clear (4) / moderately clear (3) / slightly clear (2) / confusing (1)**. |
| Relevance-IT/IS | The gathered evidence is **exceptionally outstanding (6) / very strong (5) / strong (4) / moderately compelling (3) / slightly weak (2) / weak (1)** that the proposal proves useful for the company. |
| Relevance-CS/DAA | The proposal **is exceptionally outstanding (6) / is very likely (5) / is likely (4) / will probably (3) / is unlikely (2) / is very unlikely (1)** that it would contribute to the academic/ scientific community. |
| Method | The method used is **highly appropriate and innovative (6) / highly appropriate (5) / appropriate (4) / moderately appropriate (3) / slightly appropriate (2) / inappropriate (1)**. |
| Paper | **Exemplary**! The paper surpassed expectations (6). A comprehensive paper is **highly evident (5) / evident (4) / moderately evident (3) / slightly evident (2) / not evident (1)**. |
| Mastery of the Subject | The proponents expressed their responses **correctly** and **concisely** (6). The proponents answered **all (5) / most (4) / many (3) / only few (2) / none (1)** of the questions correctly. |

**Table 7. Final Version of Proposed Assessment Tool for Software Project Stage**

| CRITERIA | |
|---|---|
| Functionality | **Exemplary**! The program performed beyond the required expectations (6). / The program was able to perform **all (5) / most (4) / many (3) / few (2)** functions as specified. / The program **did not** function at all (1). |
| Completeness | **Exemplary**! The program provides other modules beyond the required expectations (6). The program contains **all (5) / most (4) / many (3) / only a portion (2)** of the required modules. The program **did not contain (1)** any of the required modules. |
| Reliability | The codes are **bug-free** and follow **coding standards** (6). The software is error-free (5). /**Errors are evident but they do not** compromise the performance of the software (4). / **Errors are evident** and they compromise the performance of the software to **some extent** (3). / **There are errors that affect** the **overall** software performance (2). / There are so many errors to the extent that the software **no longer** performs its functions (1). |
| Paper | **Exemplary**! The paper surpassed expectations. (6) /A comprehensive paper is **highly evident (5) / evident (4) / moderately evident (3) / slightly evident (2) / not evident (1)**. |
| Mastery of the Subject | The proponents expressed their responses **correctly** and **concisely** (6). The proponents answered **all (5) / most (4) / many (3) / only few (2) / none (1)** of the questions correctly. |

Faculty commented that the confidence of students in answering questions and individual performance was not reflected in the proposed rubrics. Though important, confidence was not one of the ILOs. Thus, it was not measured in the rubrics. In terms of individual performance, the eight faculty members decided to incorporate individual rating which is not part of the rubrics. It was also disclosed that the proposed rubrics had items that were non-atomic. Non-atomic items are questions that can still be broken down to two different questions or questions that refer to the same question. For example, an item under the Paper criterion states "A comprehensive, clear, and extensive research is highly evident" is a non-atomic item. The words "comprehensive" and "extensive" may mean the same thing. Further, the "clarity" of the paper was already measured under Analysis. These concerns, comments, and suggestions were all incorporated in the first versions of the rubrics. The final rubrics are shown in Table 6 and Table 7.

## 4. CONCLUSIONS AND DIRECTIONS FOR FUTURE RESEARCH

This study attempted to develop valid and relevant capstone project courses rubrics. It was shown that it was possible to meet this goal provided that the faculty and students were consulted. As such, the final versions of the rubrics will be used for the upcoming oral defense of this current semester. Nonetheless, at this point, the strengths and limitations of the rubrics are not yet known. The actual use of the rubrics during the defense day can identify the areas to be improved with the new assessment tool. It can also be noted that the College invites external evaluators who are industry practitioners. Thus, an orientation will be held prior the oral defense. This will provide not only an avenue for the external evaluators to internalize the new assessment tool but also an opportunity to comment on the tool. The comments of external evaluators and issues concerning the instrument that might be raised during the orientation and defense will be documented and incorporated to enhance the rubrics.

## 5. ACKNOWLEDGMENTS

The researcher is thankful to the valuable contributions of the faculty members and students involved in the study. This study is made possible through the generous funding of the University of the East.